\documentclass[aps,prl,twocolumn,superscriptaddress,amsmath,amssymb,showpacs,nofootinbib]{revtex4}

\usepackage{graphicx}
\usepackage{dcolumn}
\usepackage{bm}
\usepackage{color}

\begin{document}

\title{Topological Superconductivity in Cu$_x$Bi$_2$Se$_3$}

\author{Satoshi Sasaki}
\author{M. Kriener}
\author{Kouji Segawa}
\affiliation{Institute of Scientific and Industrial Research,
Osaka University, Ibaraki, Osaka 567-0047, Japan}

\author{Keiji Yada}
\author{Yukio Tanaka}
\affiliation{Department of Applied Physics, Nagoya University, 
Nagoya 464-8603, Japan}

\author{Masatoshi Sato}
\affiliation{Instituite for Solid State Physics, University of Tokyo, 
Chiba 277-8581, Japan}

\author{Yoichi Ando}
\email{y_ando@sanken.osaka-u.ac.jp}
\affiliation{Institute of Scientific and Industrial Research,
Osaka University, Ibaraki, Osaka 567-0047, Japan}

\date{\today}

\begin{abstract}

A topological superconductor (TSC) is characterized by the topologically-protected 
gapless surface state that is essentially an Andreev bound state consisting of 
Majorana fermions.  While a TSC has not yet been discovered, the doped topological 
insulator Cu$_x$Bi$_2$Se$_3$, which superconducts below $\sim$3 K, has been predicted to 
possess a topological superconducting state.  We report that the point-contact spectra 
on the cleaved surface of superconducting Cu$_x$Bi$_2$Se$_3$ present a  
zero-bias conductance peak (ZBCP) which signifies unconventional superconductivity.  
Theoretical considerations of all possible superconducting states help us conclude 
that this ZBCP is due to Majorana Fermions and gives evidence for a topological 
superconductivity in Cu$_x$Bi$_2$Se$_3$.  In addition, we found an unusual pseudogap 
that develops below $\sim$20 K and coexists with the topological superconducting state.

\end{abstract}



\maketitle

The recent discovery of the topological insulator
\cite{KM05,BHZ06,Mol07,K1,MB,Roy,SCZ1,K2,H1,Taskin,Matsuda,Shen1,H4,H5,Sato,
Kuroda,Shen2,BTS_Rapid,BTS_Hasan,Xiong,BSTS,HK_RMP10,M_N10,Wang} stimulated
the search for an even more exotic state of matter, the topological
superconductor (TSC) \cite{QZ,Schny_B08,Sato_B10,FK_L08}. A topological
state of matter is characterized by a topological structure of the
quantum-mechanical wavefunction in the Hilbert space. In topological
insulators, a non-trivial Z$_2$ topology of the bulk valence band leads
to the emergence of Dirac fermions on the surface \cite{HK_RMP10,M_N10}.
Similarly, in TSCs non-trivial Z or Z$_2$ topologies of the
superconducting (SC) states lead to the appearance of Majorana fermions
on the surface \cite{QZ,Schny_B08,Sato_B10}. Majorana fermions are
peculiar in that particles are their own antiparticles, and they were
originally conceived as mysterious neutrinos \cite{Wil_NP09}. Currently
their realization in condensed matter is of significant interest because
of their novelty as well as the potential for quantum computing
\cite{Wil_NP09}.

The Cu$_x$Bi$_2$Se$_3$ superconductor \cite{Hor_L10,Wray_NP10,MKR_L11,MKR2}
is a prime candidate of TSC because of its peculiar band structure and
strong spin-orbit coupling \cite{FB_L10}. In this material,
Cu atoms are intercalated into the layered topological insulator
Bi$_2$Se$_3$ and the SC state appears for the Cu concentration $x$ of
about 0.2 -- 0.5, which causes electron doping with the density of
$\sim$ 10$^{20}$ cm$^{-3}$. This material has not been well studied
because of the difficulty in preparing high-quality samples
\cite{Hor_L10,Wray_NP10} but a recent breakthrough in the synthesis of
Cu$_x$Bi$_2$Se$_3$ by using electrochemistry \cite{MKR_L11,MKR2} made it
possible to prepare reliable junctions and perform a conductance
spectroscopy in the superconducting state.

\begin{figure}
\includegraphics*[width=8.7cm]{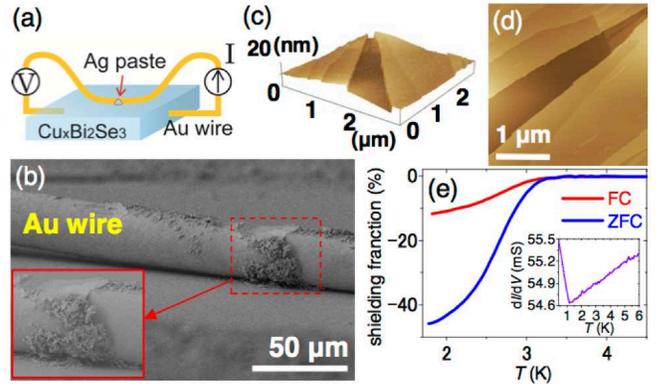}
\caption{(Color online) 
Point-contact experiment and the sample. 
(a) Sketch of the ``soft" point contact and the measurement circuit.
(b) Scanning-electron-microscope picture of the actual sample; inset magnifies 
the silver-paste spot where the point contact is formed.
(c) 3D presentation of nanometer-scale terraces on a typical cleaved surface of 
Cu$_x$Bi$_2$Se$_3$ seen by an atomic-force microscope. Typical terrace width is 0.5 $\mu$m.
(d) A false color mapping of (c).
(e) SQUID data for the SC transition in the sample ($x$ = 0.3) used for the 
point-contact measurements shown in Fig. 2. Both the zero-field-cooled (ZFC) and 
the field-cooled (FC) data measured in 0.2 mT are shown, and the former gives the SC 
shielding fraction of 46\%. Inset shows the temperature dependence of the zero-bias 
differential conductance of the point contact reported in Fig. 2.
} 
\label{fig1}
\end{figure}

\begin{figure*}
\includegraphics*[width=12cm]{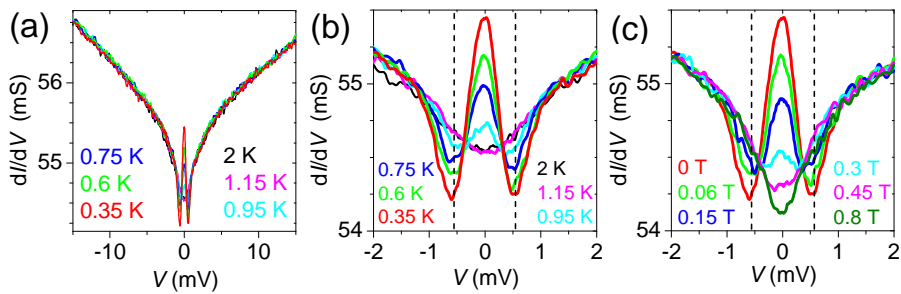}
\caption{(Color online) 
Zero-bias conductance peak. 
(a) Point-contact spectra ($dI/dV$ vs bias voltage) of Cu$_x$Bi$_2$Se$_3$ with 
$x = 0.3$ for 0.35--2 K measured in 0 T for a wide energy window. 
(b) A narrower window of (a).  
(c) The spectra at 0.35 K measured in perpendicular magnetic fields of 0--0.8 T.  
The vertical dashed lines in (b) and (c) indicate the energy position of the dips. 
} 
\label{fig2}
\end{figure*}

In the present work, we employed the so-called ``soft" point-contact
technique \cite{DG_SST10}: The contacts were prepared at room
temperature in ambient atmosphere by putting a tiny ($\sim$20 $\mu$m)
drop of silver paste on the cleaved (111) surface of a
Cu$_x$Bi$_2$Se$_3$ single crystal below a 30-$\mu$m-diameter gold wire
[Figs. 1(a) and 1(b)]. In this type of junctions, ballistic transport
occurs sporadically through parallel nanometer-scale channels formed
between individual grains in the silver paste and the sample surface
[see Figs. 1(c), 1(d) and Ref. \onlinecite{SM}]. The $dI/dV$ spectra were measured
with a lock-in technique by sweeping a dc current that is superimposed
with a small-amplitude ac current [1.35 $\mu$A (rms), corresponding to 0.5
A/cm$^2$]. We used a quasi-four-probe configuration, in
which the current was applied between a contact pad and the gold wire,
and the voltage between the wire and another contact pad was measured
[Fig. 1(a)]. The Quantum Design PPMS was used for cooling the samples
down to 0.35 K and applying the magnetic field up to 9 T.

A set of point-contact data taken on a Cu$_x$Bi$_2$Se$_3$ sample with
the bulk onset $T_\textrm{c}$ = 3.2 K is shown in Fig. 2, where one can
see that a pronounced zero-bias conductance peak (ZBCP) develops at low
temperature \cite{SM}. The inset of Fig. 1(e) shows the temperature
dependence of the zero-bias conductance, which indicates that this peak
appears below 1.2 K \cite{SM}. We note that essentially the same ZBCP 
data have been obtained on another sample (see Fig. S2 of Ref. \onlinecite{SM}).

Since heating effects can cause a spurious ZBCP \cite{Sheet_B04}, it is
important to elucidate that it is not the case here. It was argued by
Sheet {\it et al.} \cite{Sheet_B04} that in samples with a large
normal-state resistivity when the point contact is in the thermal
regime, a spurious ZBCP could show up if the increase in the bias
voltage causes the local current to exceed the critical current, which
leads to a voltage-dependent decrease in the differential conductivity.
If this is the case, the conductivity at zero bias (which is always
measured below the critical current) should {\it not} change with a weak
magnetic field; the role of the magnetic field in this case is primarily
to reduce the critical current, so the width of the spurious ZBCP would
become narrower, but the height at $V$ = 0 should be mostly unchanged as
long as the superconductor is in the zero-resistivity state. In the
magnetic-field dependence of our spectra shown in Fig. 2(c), by
contrast, the ZBCP is strongly suppressed with a modest magnetic field
while its width is little affected, which clearly speaks against the
heating origin of the ZBCP. (The magnetic field was applied perpendicular
to the cleaved surface.) Another well-known signature of the heating
effect is a sharp, spike-like dip at energies much larger than the gap
\cite{DG_SST10,Sheet_B04}, which is caused by the local transition to
normal state; in fact, when we made the point contact on a disordered
surface, we observed a widening of the peak and a lot of sharp dips at
relatively high energies, which are obviously caused by the heating
\cite{SM}. In contrast, the data shown in Fig. 2 are free from such
features, which corroborates the intrinsic nature of the ZBCP.
Therefore, one can safely conclude that the ZBCP observed here is not
due to the heating effects and is intrinsic.

One should also keep in mind that, even when the ZBCP is intrinsic, it
can be caused by several mechanisms in point contacts \cite{KT_R00}:
conventional Andreev reflection \cite{BTK_B82,Deutscher}, reflectionless
tunneling \cite{Kastalsky,Beenakker,vanWees}, magnetic scattering
\cite{Appelbaum,Rowell}, and the unconventional Andreev bound state
(ABS) \cite{KT_R00,Deutscher}. In this respect, it is important to
notice that the ZBCP shown in Fig. 2 is accompanied by pronounced dips
on its sides and the peak does not split into two even at the lowest
temperature (0.35 K). These features are clearly at odds with the
Blonder-Tinkham-Klapwijk (BTK) theory for conventional Andreev
reflection \cite{BTK_B82}. Also, the reflectionless tunneling and the
magnetic scattering are obviously irrelevant, because the former is
suppressed by a very small magnetic field of less than 0.1 T \cite{note1} 
and the
latter presents a peak splitting in magnetic fields \cite{note2}. 
Hence, one can
conclude that the ZBCP observed here is a manifestation of the ABS
\cite{KT_R00}.

Previously, it was inferred \cite{MKR_L11} from the specific-heat data
that the superconducting gap of Cu$_x$Bi$_2$Se$_3$ at $T = 0$ K, $\Delta
(0)$, would be about 0.7 meV. In Fig. 2, one can see that the minima in
the pronounced dips are located at $\sim\pm 0.6$ meV at 0.35 K; since
the ZBCP due to the ABS is usually accompanied by dips near the gap
energy \cite{KT_R00}, the energy scale of the dip is assuring.

\begin{figure*}
\includegraphics*[width=12cm]{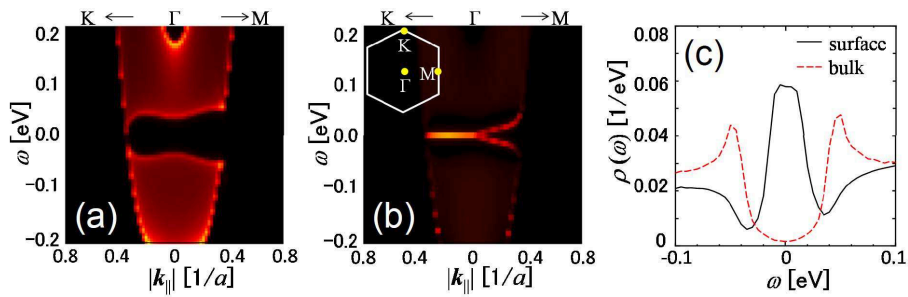}
\caption{(Color online) 
Model calculations of the topological band structure in the
superconducting state of Cu$_x$Bi$_2$Se$_3$. Theoretically
calculated spectral functions $A(\textbf{k},\omega)$ of the bulk (a) and
the surface on the $xy$ plane (b) in $\Gamma$--M and $\Gamma$--K
directions in the surface Brillouin zone shown in the inset of (b), as
well as the LDOS (c), in the superconducting state for the topological
gap function $\Delta _4$ ($\Delta _{\uparrow \uparrow} ^{12} =
\Delta_{\downarrow \downarrow }^{12} = $ $-\Delta_{\uparrow
\uparrow}^{21} = -\Delta_{\downarrow \downarrow}^{21}$); the model
Hamiltonian and the band parameters used are described in detail in 
Ref. \onlinecite{SM}. The false colour mappings of
$A(\textbf{k},\omega)$ in (a) and (b) are in arbitrary units. $\Delta (0)$
was set to be 0.05 eV for the convenience of the calculations.
} 
\label{fig3}
\end{figure*}

Given that the observed ZBCP is intrinsic and is due to the ABS, it is
important to understand its concrete origin. The ABS is caused by the
interference of the SC wavefunction at the surface, and it is a
signature of unconventional superconductivity \cite{KT_R00}. Its
occurrence is determined by the symmetry of the SC state, which in turn
is determined by the symmetry of the Hamiltonian and the pairing
mechanism. Also, it has been elucidated that Majorana fermions reside in
an ABS when it is spin non-degenerate \cite{Linder_L10}. Hence, we
examined all possible SC states in Cu$_x$Bi$_2$Se$_3$ and the nature of
the ABS to elucidate whether the observed ZBCP is due to Majorana
fermions. The microscopic model to describe the band structure of
Cu$_x$Bi$_2$Se$_3$ has already been developed 
\cite{FB_L10,hamiltonian,hamiltonian2,HaoLee_B11}, 
and it was shown \cite{FB_L10} that, if both short- and long-range
interactions are considered, the symmetry of the Hamiltonian allows four
different types of the SC gap function, $\Delta _1$ to $\Delta _4$
\cite{SM}, with three of them being unconventional. Following Ref. 
\onlinecite{HaoLee_B11}, we
have theoretically calculated the spectral functions of the bulk and the
surface as well as the local density of states (LDOS) for all possible
gap functions (see Ref. \onlinecite{SM} for details), similar to those
done in Refs. \onlinecite{Schny_B10} and \onlinecite{Yada}.

Firstly, the conventional even-parity SC state $\Delta _1$ was found to
give no two-dimensional (2D) ABS \cite{SM}. While in this case the
surface could become a 2D TSC due to the proximity effect as proposed by
Fu and Kane \cite{FK_L08}, the surface of a three-dimensional (3D)
superconductor is continuously connected and has no topological edge;
hence, the one-dimensional Majorana fermions that might appear at the
edge of a 2D TSC \cite{FK_L08} would not exist in the present case.

Among the remaining three possible SC states that are all
unconventional, the fully-gapped, odd-parity SC state $\Delta _2$ gives
rise to 2D helical Majorana fermions as the ABS. However, because of the
Dirac-like dispersion of this ABS, the surface LDOS tends to have a
minimum at zero energy \cite{SM}, which does not agree well with our
data; nevertheless, it was very recently proposed that the ZBCP could
appear even in this fully-gapped state due to a peculiar ``twisting" of
the ABS dispersion \cite{Hsieh-Fu}. In the case of the other two
odd-parity SC states, $\Delta _3$ and $\Delta _4$, both of which have
two point nodes, a single ZBCP naturally shows up in the surface LDOS
[Figs. 3(a) -- 3(c)]; this is because the point nodes lead to a
partially flat dispersion of the helical Majorana fermions,
concentrating the LDOS near zero energy. Therefore, it is most likely
that the observed ZBCP signifies 2D Majorana fermions due to the
odd-parity bulk SC state, although it is difficult to determine the
exact pairing state from the three possibilities at this stage. The fact
that the ZBCP is strongly suppressed with a modest magnetic field [Fig.
2(c)] supports this conclusion, because the helical Majorana fermions
are naturally suppressed as the time-reversal symmetry is broken with
the magnetic field. Note that, while there are nanometer-scale terraces
on the cleaved surface [Figs. 1(c), 1(d) and Ref. \onlinecite{SM}],
electron transmissions in the in-plane directions through the side walls
of the terraces are much less likely to take place compared to the
transmissions in the out-of-plane direction, because the typical terrace
height ($<$ 10 nm) is much smaller than the typical Ag grain size of 50 nm
\cite{SM}. Therefore, our data are expected to reflect mostly the ABS on
the (111) surface.

We now discuss the topological nature of the possible SC states $\Delta _3$ and
$\Delta _4$. The presence of the point nodes might seem to preclude the
topological superconductivity, which is usually considered to require a
full gap. However, for the $\Delta _3$ and $\Delta _4$ states one can
define a non-trivial topological invariant, ``mod-2 winding number",
which is immune to weak perturbations and assures that the $\Delta _3$
and $\Delta _4$ states are robustly topological \cite{SM}. In fact, a
time-reversal-invariant SC state with a pair of point nodes is
adiabatically connected to a fully-gapped state in the ``mod-2 winding
number" topological class, and having an odd parity is sufficient for
this case to become topologically non-trivial \cite{SM}.

Previously, we reported that the specific-heat data 
was most consistent with a fully-gapped SC state \cite{MKR_L11}. It is fair to note, 
however, that the entropy 
contribution of the quasiparticles excited near the point node of a 3D SC state is very 
small and, indeed, the $T^3$ dependence of the specific heat expected for point nodes 
is difficult to be distinguished \cite{Ott_L84}, particularly in inhomogeneous samples.  
Therefore, the $\Delta _3$ or $\Delta _4$ state with point nodes does not necessarily 
contradict the existing specific-heat data.

An interesting and unexpected feature in our data is that a pseudogap develops below 
$\sim$20 K [Fig. 4(a)].  As shown in Figs. 4(b) -- 4(g), this pseudogap appears to be 
enhanced by the magnetic field, and it is most pronounced at 0.35 K in high magnetic 
fields.  This pseudogap coexists with the superconductivity below the upper critical 
field $H_{c2}$ \cite{note3} and may give us a clue to 
understanding the paring mechanism in Cu$_x$Bi$_2$Se$_3$. Finally, how the spectra 
change with temperature and magnetic field is summarized in false colour mappings shown 
in Figs. 4(h) and 4(i).

\begin{figure}
\includegraphics*[width=8.7cm]{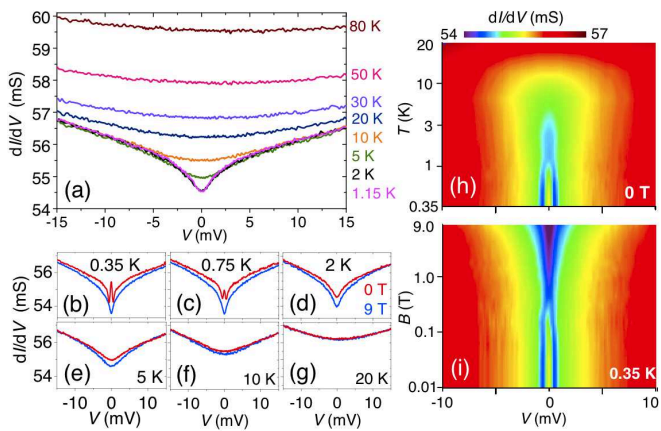}
\caption{(Color online) 
Pseudogap in Cu$_x$Bi$_2$Se$_3$.
(a) $dI/dV$ vs bias voltage for 1.15--80 K measured in 0 T. (b -- g)
Comparisons of the spectra in 0 and 9 T. At low temperature below $\sim
$20 K, the spectra in 9 T show smaller $dI/dV$ near zero bias compared
to that at 0 T, indicating that the pseudogap deepens with the magnetic
field. (h, i) False colour mappings of $dI/dV$ in the bias-voltage
vs temperature plane in 0 T (h) and in the bias-voltage vs
magnetic-field plane at 0.35 K (i), summarizing how the spectra change
with temperature and magnetic field; note that the vertical axes are in
logarithmic scales in both (h) and (i).
} 
\label{fig4}
\end{figure}

As is clear from the above discussions, one can conclude that the ZBCP
in Cu$_x$Bi$_2$Se$_3$ signifies an ABS consisting of 2D Majorana
fermions and that Cu$_x$Bi$_2$Se$_3$ is hosting a topological
superconductivity. It is therefore an urgent task to determine the exact
pairing symmetry in Cu$_x$Bi$_2$Se$_3$. Regarding the Majorana physics,
an interesting question is the existence of the Majorana zero-mode in
the vortex core \cite{Hosur}. The 2D Majorana fermions living on the surface of a 3D
TSC are different from the non-Abelian Majorana fermions of a 2D TSC
proposed for topological quantum computing \cite{HK_RMP10, FK_L08}, but
establishing a general understanding of Majorana fermions is important
for both fundamental physics and future information technologies.

In summary, our point-contact spectroscopy of the Cu$_x$Bi$_2$Se$_3$
superconductor found an unusual pseudogap below $\sim$20 K and a
pronounced ZBCP in the SC state. The latter signifies an unconventional
SC state that can only be topological in Cu$_x$Bi$_2$Se$_3$, and
therefore our observation gives evidence for a topological
superconductivity in this material. One can fully expect that
Cu$_x$Bi$_2$Se$_3$ as the first concrete example of a TSC will greatly
help advance our understanding of topological states of matter and
associated exotic quasiparticles.

\begin{acknowledgments} 

We thank L. Fu, A. Furusaki and S. Onoda for useful discussions, and K.
Matsumoto and S. Wada for their help in the experiment. This work was
supported by JSPS (NEXT Program), MEXT (Innovative Area ``Topological
Quantum Phenomena" KAKENHI), and AFOSR (AOARD 10-4103).

\end{acknowledgments}

\clearpage
\onecolumngrid

\renewcommand{\thefigure}{S\arabic{figure}} 

\setcounter{figure}{0}

\renewcommand{\thesection}{S\arabic{section}.} 

\begin{flushleft} 
{\Large {\bf Supplemental Material}}
\end{flushleft} 

\vspace{2mm}

\begin{flushleft} 
{\bf S1. Surface of Cu$_x$Bi$_2$Se$_3$ single-crystal samples}
\end{flushleft} 

Although the cleaved surface of Cu$_x$Bi$_2$Se$_3$ single crystals is
essentially flat [Fig. S1(a)], there are a lot of nanometer-scale
terraces as revealed by an atomic-force microscope [Figs. S1(b) and
S1(c), and also Figs. 1(c) and 1(d) of the main text]. The heights of
those terraces are often larger than the one quintuple-layer unit (0.95
nm) but are typically less than 10 nm. There is no clear preferential
direction for the edges of the terraces. While there are several tens of
terraces in the lateral area of a 20-$\mu$m point contact in which
nanometer-scale silver grains (typically 50-nm diameter) are
distributed, electron transmissions into the Ag grains from the (111)
surface is expected to be dominant because the typical terrace height ($<$ 10 nm) 
is much smaller than the typical Ag grain size (50 nm).

\begin{flushleft} 
{\bf S2. Reproducibility of the ZBCP}
\end{flushleft} 

To demonstrate the reproducibility of the observed ZBCP, Figs. S2(a) and S2(b) 
present a set of point-contact spectra showing the ZBCP below 1 K,
taken on another Cu$_x$Bi$_2$Se$_3$ sample with the bulk onset $T_c$ =
3.2 K. At 0.35 K, the minima in the dips are located at about $\pm$0.4
meV, which suggests that the SC gap in this sample was a bit smaller compared 
to the sample shown in the main text.

We note that the chances of observing the ZBCP
were not very high: We have so far measured 47 samples, and 25
of them showed no ZBCP down to 0.35 K
(an example is shown in Fig. S3). 
In other words, 22 out of 47 samples have been found to present a ZBCP
in our experiment.
This seems to be correlated with the fact
that our samples are inhomogeneous and show the SC shielding 
fractions of around 40\%.

\begin{flushleft} 
{\bf S3. Onset temperature of the ZBCP}
\end{flushleft}

As mentioned in the main text, the enhancement of the zero-bias
conductance of the sample reported in Fig. 2 occurred below 1.2 K, which
is lower than the bulk onset $T_c$ of 3.2 K. One possibility for this
difference is that the ZBCP is prone to thermal smearing and needs a low
temperature to become observable. Another possibility is that the $T_c$
was locally 1.2 K at the position beneath the point contact; this is
conceivable because the temperature dependence of the diamagnetic signal
[Fig. 1(e)] suggests a broad distribution of local $T_c$. If this second
possibility is actually the case, our observation that the energy scale
of the dip in the point-contact spectra [0.6 meV, see Fig. 2(b)] agrees
with the estimated bulk SC gap may suggest an interesting situation
where the pair potential $\Delta(0)$ is essentially the same in all the
superconducting portions of the sample but the local $T_c$ is determined
by the local carrier density, because the carrier density in
Cu$_x$Bi$_2$Se$_3$ is very low and the $T_c$ may well be governed by the
superfluid density \cite{Kivelson}.

\begin{flushleft} 
{\bf S4. Effects of heating and/or critical currents on the point-contact spectra}
\end{flushleft} 

When the point contacts were made on disordered surfaces, such as the
as-prepared surface after the electrochemical reaction [Figs. S4(c) and
S4(d)], 
we observed a widening of the central peak and a lot 
of sharp dips at the tail, as shown in Figs. S4(a) and S4(b). 
Those sharp dips have been discussed \cite{Sheet_B04} to be due to the effects of 
heating and/or critical currents. Although the spectra in Fig. S4 are
contaminated by the heating effects, it should be noted that 
the sharp central part within the two vertical dashed lines in Fig. S4(b)
is likely to signify the intrinsic ZBCP, because its width is less
than 2 meV and is consistent with twice the gap energy.


\begin{flushleft} 
{\bf S5. Theoretical calculations of the surface states of Cu$_x$Bi$_2$Se$_3$}
\end{flushleft} 

\setcounter{MaxMatrixCols}{10}

It is known from the studies of the tunneling spectroscopy of 
unconventional superconductors \cite{KT_R00} that
the measured conductance of actual experiments of a point contact
between a normal metal and a superconductor 
corresponds to the local density of states (LDOS) of the superconductor
at the surface.
Therefore, we theoretically calculate the LDOS of Cu$_x$Bi$_2$Se$_3$ 
based on a model Hamiltonian for the topological insulator 
Bi$_2$Se$_3$ proposed by H. Zhang {\it et al.} \cite{hamiltonian}. 
The Hamiltonian ${\cal H}({\bm k})$ of the present system becomes 
8 $\times$ 8 matrix, due to the 
presence of two orbitals, spin indices and 
electron-hole space \cite{FB_L10, HaoLee_B11}. 
We denote ${\cal H}({\bm k})$ using the $4 \times 4$ matrices $\hat \xi({\bm k})$ 
and $\hat{\Delta}$ as 
\begin{equation}
{\cal H}({\bm k}) =
\left(
\begin{array}{cc}
\hat \xi({\bm k}) & \hat{\Delta} \\
\hat{\Delta}^{\dagger} & -\hat \xi^{*}(-{\bm k})
\end{array}
\right).
\end{equation}
The normal-state Hamiltonian $\hat \xi({\bm k})$ is given by
\begin{eqnarray}
\hat \xi({\bm k})=
\left(
\begin{array}{cccc}
\varepsilon({\bm k})+M({\bm k}) & 0 & A_1({\bm k}) & A_2^-({\bm k})\\
0 & \varepsilon({\bm k})+M({\bm k}) & A_2^+({\bm k}) & -A_1({\bm k})\\
A_1({\bm k}) & A_2^-({\bm k}) & \varepsilon({\bm k})-M({\bm k}) & 0\\
A_2^+({\bm k}) & -A_1({\bm k}) & 0 & \varepsilon({\bm k})-M({\bm k})
\end{array}
\right)
\label{hamiltonian}
\end{eqnarray}
with
\begin{eqnarray}
\varepsilon({\bm k})&=&\bar D_1(2-2\cos \left(k_zc\right))\\ \nonumber
&+&\frac{4}{3}\bar D_2\left(3-2\cos\left(\frac{\sqrt3}{2}k_xa\right)\cos\left(\frac{1}{2}k_ya\right)-\cos \left(k_ya\right)\right)-\mu\\
A_1({\bm k})&=& \bar A_1\sin \left(k_zc\right)\\
A_2^\pm({\bm k})&=&\frac{2}{3}\bar{A}_{2} \\ \nonumber
&\times& \left\{\sqrt3\sin\left(\frac{\sqrt3}{2}k_xa\right)\cos\left(\frac{1}{2}k_ya\right)\pm i\left(\cos\left(\frac{\sqrt3}{2}k_xa\right)\sin\left(\frac{1}{2}k_ya\right)+\sin \left(k_ya\right)\right)\right\}\\ 
M({\bm k})&=&M_0-\bar B_1(2-2\cos \left(k_zc\right))\\
&-&\frac{4}{3}\bar B_2\left(3-2\cos\left(\frac{\sqrt3}{2}k_xa\right)\cos\left(\frac{1}{2}k_ya\right)-\cos \left(k_ya\right)\right). \nonumber
\end{eqnarray}
Note that in the present calculations, we consider the hexagonal lattice 
with the lattice constants $a$ and $c$, where 
the 2D triangular lattices stack along the $c$-axis direction.
We use the same values of parameters 
$M_0$, $\bar A_2$, $\bar B_2$ and $\bar D_2$ 
as given in Ref. \cite{hamiltonian},
with the transformation 
$\bar A_2=A_2/a$, $\bar B_2=B_2/a^2$ and $\bar D_2=D_2/a^2$. 
The values of $\bar A_1$, $\bar B_1$, and $\bar D_1$ are chosen as 
$\bar A_1=0.32$ eV, $\bar B_1=0.216$ eV, and $\bar D_1=0.024$ eV 
to fit the energy dispersion 
for the $\Gamma-Z$ direction obtained in Ref. \cite{hamiltonian}. 
The chemical potential $\mu$, which is measured from the Dirac point, was 
estimated form the experimental data \cite{Wray_NP10} to be $\mu=0.5$ eV. 

The $4 \times 4$ pair-potential matrix $\hat{\Delta}$ is expressed as
\begin{eqnarray}
\hat{\Delta}=
\left(
\begin{array}{cccc}
\Delta^{11}_{\uparrow\uparrow} & \Delta^{11}_{\uparrow\downarrow} & \Delta^{12}_{\uparrow\uparrow} & \Delta^{12}_{\uparrow\downarrow}\\
\Delta^{11}_{\downarrow\uparrow} & \Delta^{11}_{\downarrow\downarrow} & \Delta^{12}_{\downarrow\uparrow} & \Delta^{12}_{\downarrow\downarrow}\\
\Delta^{21}_{\uparrow\uparrow} & \Delta^{21}_{\uparrow\downarrow} & \Delta^{22}_{\uparrow\uparrow} & \Delta^{22}_{\uparrow\downarrow}\\
\Delta^{21}_{\downarrow\uparrow} & \Delta^{21}_{\downarrow\downarrow} & \Delta^{22}_{\downarrow\uparrow} & \Delta^{22}_{\downarrow\downarrow}
\end{array}
\right).
\end{eqnarray}
Upon treating $\hat{\Delta}$,
we neglect the ${\bm k}$-dependence for simplicity. 
Following Hao and Lee \cite{HaoLee_B11}, we consider the four types of $\hat{\Delta}$ 
as shown in Table 1. Note that in Ref. \cite{HaoLee_B11} the total number of possible 
$\hat{\Delta}$ are six; however, two of them have their counterpart which is
the same gap type, so the four 
types of $\hat{\Delta}$ summarized in Table 1 are exhaustive.
In all the gap types, $\Delta^{lm}_{\sigma\sigma'}=-\Delta^{ml}_{\sigma'\sigma}$ 
is satisfied in accordance with  the Fermi-Dirac statistics.

\begin{table}[t]
\begin{center}
\begin{tabular}{|c|c|c|c|}
\hline
&gap type& parity & energy gap structure \\ \hline\hline
$\Delta_1$&
$\Delta^{11}_{\uparrow\downarrow}=-\Delta^{11}_{\downarrow\uparrow}=
\Delta^{22}_{\uparrow\downarrow}=-\Delta^{22}_{\downarrow\uparrow}$ &
even & full gap\\
& $\Delta^{11}_{\uparrow\downarrow}=-\Delta^{11}_{\downarrow\uparrow}=
-\Delta^{22}_{\uparrow\downarrow}=\Delta^{22}_{\downarrow\uparrow}$ &
& \\ \hline
$\Delta_2$&
$\Delta^{12}_{\uparrow\downarrow}=-\Delta^{12}_{\downarrow\uparrow}=
\Delta^{21}_{\uparrow\downarrow}=-\Delta^{21}_{\downarrow\uparrow}$ &
odd & full gap \\ \hline
$\Delta_3$&
$\Delta^{12}_{\uparrow\downarrow}=\Delta^{12}_{\downarrow\uparrow}=
-\Delta^{21}_{\uparrow\downarrow}=-\Delta^{21}_{\downarrow\uparrow}$ &
odd & point node\\ \hline
$\Delta_4$&
$\Delta^{12}_{\uparrow\uparrow}=\Delta^{12}_{\downarrow\downarrow}=
-\Delta^{21}_{\uparrow\uparrow}=-\Delta^{21}_{\downarrow\downarrow}$ &
odd & point node\\
& $\Delta^{12}_{\uparrow\uparrow}=-\Delta^{12}_{\downarrow\downarrow}=
-\Delta^{21}_{\uparrow\uparrow}=\Delta^{21}_{\downarrow\downarrow}$ &
& \\
\hline
\end{tabular}
\end{center}
\caption{Four types of pair potentials introduced by Hao and Lee \cite{HaoLee_B11}.
Odd-parity gap functions correspond to unconventional SC states.}
\end{table}

To obtain the LDOS, we used the same calculation technique as that 
employed in Ref. \cite{Yada}; namely, we calculate the Green's function 
for the (001) flat surface [which corresponds to the (111) surface in the 
rhombohedral notation] without disorder,  
where the momentum parallel to the surface ${\bm k}_\parallel=(k_x, k_y)$ 
is conserved. 
The surface Green's function is constructed by introducing an 
infinite potential barrier at $z=z_0$; 
with a sufficiently large system size along the $z$-direction,
one can identify the surface states as the states at $z=z_1$ which is 
next to the infinite potential barrier in our lattice calculations.
The surface Green's function $\check G_s({\bm k}_\parallel,\omega)$ 
is obtained from the Fourier-transformed form of the bulk Green's function
$\check G_b(z-z';{\bm k}_\parallel, \omega)$ which is written as
\begin{equation}
\check G_b(z-z';{\bm k}_\parallel, \omega)
=\frac{1}{N_z}\sum_{k_z}
\check G_b(\bm{k}_\parallel,k_z,\omega)e^{ik_z(z-z')},
\end{equation}
with 
\begin{equation}
\check G_b(\bm{k}_\parallel,k_z,\omega)
= \frac{1}{\omega - {\cal H}(\bm {k})},
\end{equation}
and $\check G_s({\bm k}_\parallel,\omega)$ is given by
\begin{eqnarray}
\check G_s({\bm k}_\parallel,\omega)&=&\check G_b(z_1-z_1;k_y,\omega)\nonumber\\&-&\check G_b(z_1-z_0;k_y,\omega)\{\check G_b(z_0-z_0;k_y,\omega)\}^{-1}\check G_b(z_0-z_1;k_y,\omega), 
\end{eqnarray}
based on the T-matrix method.
The resulting  
momentum-resolved spectral function $A({\bm k}_\parallel,\omega)$ is given by
\begin{eqnarray}
A({\bm k}_\parallel,\omega)=-\frac{1}{\pi}\sum_{\alpha=1-4}{\rm Im} \{\check G_s({\bm k}_\parallel,\omega)\}_{\alpha\alpha}. 
\end{eqnarray}
Finally, we obtain the LDOS $\rho_s(\omega)$ by integrating 
$A({\bm k}_\parallel,\omega)$ over momentum, 
\begin{eqnarray}
\rho_s(\omega)=\frac{1}{N_xN_y}\sum_{{\bm k}_\parallel}A({\bm k}_\parallel,\omega). 
\end{eqnarray}
In the actual calculations, we took the mesh size of $N_x=N_y=516$ and 
$N_z=512$, and used the maximum gap size, $\Delta$(0), of 0.05 eV for 
convenience.

Let us discuss the calculated $A({\bm k}_\parallel,\omega)$ shown in
Figs. S5 to S8 for the four gap types. In those figures, $A({\bm
k}_\parallel,\omega)$ is shown by using false colour mapping in arbitrary
units along $\Gamma-$M and $\Gamma-$K directions. For $\Delta_1$ and
$\Delta_2$, the excitation spectra have a gap in the bulk [Figs. S5(a) 
and S6(a)]. At the surface, there is no surface Andreev bound
state (ABS) for $\Delta_1$ [Fig. S5(b)], but the ABS appears as a
helical edge mode for $\Delta_2$ [Fig. S6(b)]. For $\Delta_3$ and
$\Delta_4$, the bulk gap closes at certain points on the Fermi surface;
namely, those gaps have point nodes [Figs. S7(a) and S8(a)].
Closer examinations of those gap functions find that each has two point
nodes at opposing points on the Fermi surface. At the surface for
$\Delta_4$, one finds an ABS with a partially flat dispersion [Fig.
S8(b)]; such a dispersion relation in the ABS can induce a large
density of zero-energy states after integrating over $k_{\parallel}$
\cite{Tanaka}. In contrast, no clear ABS is visible at the surface for $\Delta_3$ in this
calculation [Fig. S7(b)]; however, this does not necessarily mean 
that the ABS is absent for $\Delta_3$. In fact, this is probably because the 
ABS overlaps with the bulk spectral function (whose contribution is 
also present at the surface, $z=z_1$), which could make the ABS invisible. 
(As we discuss in the next section, both the $\Delta_3$ and
$\Delta_4$ states are topological, so the bulk-edge
correspondence dictates that the ABS be present on any surface for
$\Delta_3$ and $\Delta_4$.)

In Fig. S9, the resulting LDOS, $\rho(\omega)$, is plotted for the two
unconventional cases, $\Delta_2$ [fully-gapped, Fig. S9(a)] and
$\Delta_4$ [point node, Fig. S9(b)]. The $\Delta_2$ case is similar
to the BW-phase in superfluid $^{3}$He \cite{Asano2003} and
$\rho(\omega)$ at the surface has a minimum at zero energy. On the other
hand, in the $\Delta_4$ case with point nodes, $\rho(\omega)$ has a
clear ZBCP, due to the partially-flat dispersion of the ABS.

It should be noted that the results for the $\Delta_2$ state can change
qualitatively when the parameters for the calculations are changed, and
for some range of the parameters, a ZBCP was obtained; this is
consistent with the very recent proposal by Hsieh and Fu
\cite{Hsieh-Fu}. As an example, we show in Fig. S10 the band dispersions
and the LDOS for the bulk and surface states obtained within our model
for a different set of parameters. One can see that the surface ABS has
a twisted dispersion [Fig. S10(a)] and the surface LDOS shows a
three-peak structure with the central peak at zero energy. Therefore, it
is probably premature to dismisses the fully-gapped $\Delta_2$ state as
being inconsistent with the experimental data showing a pronounced ZBCP.

\clearpage

\begin{flushleft} 
{\bf S6. Bulk topological number of the $\Delta_3$ and $\Delta_4$ states
with point nodes}
\end{flushleft} 

Generally, topological phases are characterized by nontrivial
topological numbers associated with the global structure of the Hilbert
space. Hence, one usually presumes the existence of a non-zero energy
gap which separates the topological ground state from its
excited states. However, this is not so simple if we consider
superconducting states. Indeed, in two dimensions, the existence of a
well-defined topological number has been shown even for gapless
superconductors \cite{SF10}.

Here, we generalize the arguments in Ref. \cite{SF10} and show that the
superconducting states $\Delta_3$ and $\Delta_4$, both of which have
point nodes in the gap, support a well-defined topological number. Our
finding indicates that the gapless surface states of these
superconductors are topologically protected. In other words, they are
topological superconductors.

Before examining the nodal superconductors, we briefly review the
topological number for {\it fully-gapped} 3D time-reversal-invariant
superconductors. The topological number is defined on the basis of the
symmetry of the Bogoliubov-de Gennes (BdG) Hamiltonian. A fundamental
character of the BdG Hamiltonian is the particle-hole symmetry
\begin{eqnarray}
{\cal C}{\cal H}({\bm k}){\cal C^{\dagger}}=-{\cal H}^*(-{\bm k}),
\end{eqnarray}
where ${\cal C}$ is the charge conjugation matrix. In addition, in  
time-reversal-invariant superconductors, the BdG Hamiltonian has the 
time-reversal symmetry
\begin{eqnarray}
\Theta {\cal H}({\bm k})\Theta^{\dagger}={\cal H}^*({\bm k}), 
\label{eq:time-reversal}
\end{eqnarray}
with $\Theta \Theta^T=-1$.
Combining these symmetries, one obtains the so-called chiral symmetry,
\begin{eqnarray}
\left\{{\cal H}({\bm k}), \Gamma \right\}=0, 
\quad
\Gamma=\Theta{\cal C}.
\end{eqnarray}
This chiral symmetry plays the central role in the definition of the
topological number.
Let us consider the BdG equation,
\begin{eqnarray}
{\cal H}({\bm k})|u_n({\bm k})\rangle=E_n({\bm k})|u_n({\bm k})\rangle. 
\end{eqnarray}
When the chiral symmetry is present, one obtains 
\begin{eqnarray}
{\cal H}({\bm k})\Gamma|u_n({\bm k})\rangle=-E_n({\bm k})\Gamma|u_n({\bm
 k})\rangle,
\end{eqnarray}
which means that if $|u_n({\bm k})\rangle$ is a quasiparticle
state with positive energy $E_n>0$, then $\Gamma|u_n({\bm k})\rangle$ is
a quasiparticle state with negative energy. 
We use a positive $n$ (negative $n$) for $|u_n({\bm
k})\rangle$ to represent a positive (negative)
energy quasiparticle state, and set
\begin{eqnarray}
|u_{-n}({\bm k})\rangle=\Gamma |u_n({\bm k})\rangle. 
\label{eq:negative_positive}
\end{eqnarray}
The topological number is defined by using the following $Q$ matrix, 
\begin{eqnarray}
Q({\bm k})=\sum_{n>0}|u_n({\bm k})\rangle\langle u_n({\bm k})|
-\sum_{n<0}|u_n({\bm k})\rangle\langle u_n({\bm k})|.
\end{eqnarray}
From Eq. (\ref{eq:negative_positive}), we have $\{Q({\bm k}), \Gamma\}=0$.
Hence, if we take the basis in which $\Gamma$ is diagonalized as
\begin{eqnarray}
\Gamma=\left(
\begin{array}{cc}
1 & 0\\
0 & -1
\end{array}
\right), 
\end{eqnarray}
then $Q({\bm k})$ becomes off-diagonal
\begin{eqnarray}
Q({\bm k})=\left(
\begin{array}{cc}
0 & q({\bm k})\\
q^{\dagger}({\bm k}) &0
\end{array}
\right),
\end{eqnarray}
where $q({\bm k})$ is a unitary matrix because $Q({\bm k})$ satisfies
$Q^2({\bm k})=1$. 
The topological number is
defined as the winding number of the unitary matrix $q({\bm k})$ \cite{Schny_B08}, 
\begin{eqnarray}
w_{\rm 3d}=\frac{1}{24\pi^2}\int d^3k \epsilon^{ijk}
{\rm tr}[q\partial_iq^{\dagger}q\partial_jq^{\dagger}q\partial_k q^{\dagger}]. 
\end{eqnarray}
Here, note that the above definition of the topological number is
possible only when the system is fully gapped.
Otherwise, the positive energy states can not be separated from the negative
ones, and the $Q({\bm k})$ matrix is not well defined.

Now let us try to define the same topological number for nodal
superconductors such as $\Delta_3$ and $\Delta_4$. 
The simplest way to do this is to introduce a small perturbation
to eliminate all nodes in the spectrum.
For time-reversal-invariant superconductors, point nodes are not topologically
protected, so we can always open a gap by a small deformation of the gap
functions.\footnote{Note that this is not possible when the
time-reversal-symmetry is broken. For instance, the point nodes in the ${}^3$He
A-phase cannot be removed by small perturbations, since they have their own
topological number. The time-reversal invariance is crucial in the
argument in this section.} 
After removing the nodal points,
the winding number $w_{\rm 3d}$ can be evaluated in the manner shown above.
This simple procedure, however, does not work after all. The
problem is that the value of the winding number depends on
the perturbation we choose. As a result, one cannot have a
unique definition of the winding number for gapless systems.

On the other hand, we find that this procedure does define the {\it
mod-2 winding number} ({\it i.e.}, parity of the winding number)
uniquely. The point is that we always have point nodes in pairs: From
the time-reversal invariance (or particle-hole symmetry), if we have a
point node $K$ at ${\bm k}={\bm k}_0$, then we have another node $K'$ at
${\bm k}=-{\bm k}_0$. Thus, even if the point node $K$ may cause the
ambiguity of integer $N$ in the winding number, the total ambiguity is
always $2N$. Therefore, the mod-2 winding number is not affected by
the ambiguity of this procedure.

Actually, the above statement can be shown rigorously.
For this purpose, we introduce the ``gauge field'' $A_i({\bm k})$,
\begin{eqnarray}
A_i({\bm k})_{mn}=i\langle u_m({\bm k})|\partial_i u_n({\bm k})\rangle,
\label{eq:gauge}
\end{eqnarray}
and consider the ``Chern-Simons term'',
\begin{eqnarray}
S_{\rm CS}&=&\frac{1}{16\pi^2}
\int d^3k \epsilon^{ijk} 
{\rm tr}[(F_{ij}+\frac{2}{3}iA_iA_j)A_k],
\label{eq:CS}
\end{eqnarray}
with $F_{ij}=\partial_iA_j-\partial_jA_i-i[A_i, A_j]$.
Here the matrix multiplication and the trace are done for negative $n$
({\it i.e.}, negative energy states). 
To see how the Chern-Simons term is related to the winding number, 
consider the following eigenstate of $Q({\bm k})$ with
the eigenvalue $-1$:
\begin{eqnarray}
|\psi_n({\bm k})\rangle=\frac{1}{\sqrt{2}}
\left(
\begin{array}{c}
\phi_n \\
-q^{\dagger}({\bm k})\phi_n
\end{array}
\right),
\end{eqnarray}
where $\phi_n$ is a ${\bm k}$-independent orthogonal basis vector, {\it i.e.},
$\phi_n^{\dagger}\phi_m=\delta_{mn}$.
Because a negative energy state $|u_n({\bm k})\rangle$ $(n<0)$ is also
an eigenstate of $Q({\bm k})$ with the eigenvalue $-1$,
we can  expand it with $|\psi_n({\bm k})\rangle$ as
\begin{eqnarray}
|u_n({\bm k})\rangle=\sum_m|\psi_m({\bm k})\rangle U_{mn}({\bm k}), 
\label{eq:expand}
\end{eqnarray}
with a unitary matrix $U({\bm k})$.
From the definition of the gauge field, we have 
\begin{eqnarray}
A_i({\bm k})=iU^{\dagger}\partial_i U+U^{\dagger}a_i({\bm k})U,
\quad
a_i({\bm k})=\frac{i}{2}q\partial_i q^{\dagger}. 
\label{eq:Aarelation}
\end{eqnarray}
Substituting Eq. (\ref{eq:Aarelation}) into Eq. (\ref{eq:CS}) yields
\begin{eqnarray}
S_{\rm CS}&=&\frac{1}{16\pi^2}
\int d^3k \epsilon^{ijk} 
{\rm tr}[(F_{ij}+\frac{2}{3}iA_iA_j)A_k]
\nonumber\\
&=&\frac{1}{16\pi^2}
\int d^3k \epsilon^{ijk} 
{\rm tr}[(f_{ij}+\frac{2}{3}ia_ia_j)a_k]
+\frac{1}{24\pi^2}\int d^3k
{\rm tr}[U^{\dagger}\partial_i U U^{\dagger}\partial_j U
U^{\dagger}\partial_k U]
\nonumber\\
&=&
\frac{1}{48\pi^2}\int d^3k
\epsilon^{ijk}{\rm tr}[q\partial_i q^{\dagger}q\partial_j q^{\dagger} 
q\partial_k q^{\dagger}]
+\frac{1}{24\pi^2}\int d^3k
{\rm tr}[U^{\dagger}\partial_i U U^{\dagger}\partial_j U
U^{\dagger}\partial_k U].
\label{eq:AaU}
\end{eqnarray}
Here we have used the fact that in a time-reversal-invariant system the
global basis of wave functions exists on a torus $T^3$ \cite{WQZ10}, and hence 
the total derivative term does not contribute to the integration.
Remembering that the second term in the third line of the right hand side of
Eq. (\ref{eq:AaU}) is the winding number of $U$, we obtain
\begin{eqnarray}
S_{\rm CS}=\frac{1}{2}w_{\rm 3d}+N, 
\label{eq:CSwinding}
\end{eqnarray}
where $N$ is an integer denoting the winding number of $U$.
The above relation (\ref{eq:CSwinding}) means that the mod-2
winding number is simply given by 
the parity of twice the Chern Simons term,
\begin{eqnarray}
(-1)^{w_{\rm 3d}}=(-1)^{2S_{\rm SC}}. 
\label{eq:w3dCS}
\end{eqnarray}

Now we employ a useful relation derived for topological insulators by
Wang, Qi and Zhang, who showed \cite{WQZ10} that the parity of twice the
Chern-Simons term coincides with the ${\bm Z}_2$ invariant given by Fu,
Kane and Mele \cite{FKM07} and by Moore and Balents \cite{MB07}. 
The ${\bm Z}_2$ invariant is expressed
by the Pfaffians of the following unitary matrix between the occupied
states,
\begin{eqnarray}
w_{mn}({\bm k})=\langle u_m(-{\bm k})|\Theta|u^*_n({\bm k})\rangle. 
\end{eqnarray}
Because of the identity $\Theta \Theta^T=-1$, $w_{nm}({\bm k})$ becomes
antisymmetric at the time-reversal-invariant momentum (TRIM)
$\Gamma_i$, which makes it possible to define the Pfaffians at these
points. The ${\bm Z}_2$ invariant $\nu_0$ is given by
\begin{eqnarray}
(-1)^{\nu_0}=\prod_{i=1}^{8}\delta_i, 
\quad
\delta_i=\frac{\sqrt{{\rm det}[w(\Gamma_i)]}}{{\rm Pf}[w(\Gamma_i)]},
\end{eqnarray}
where the product is taken for all the TRIMs in the Brillouin zone.
The relation given by Wang, Qi and Zhang \cite{WQZ10} is  
\begin{eqnarray}
(-1)^{2S_{\rm SC}}=(-1)^{\nu_0}. 
\label{eq:WQZ}
\end{eqnarray}
Combining Eqs. (\ref{eq:w3dCS}) and (\ref{eq:WQZ}), one obtains
\begin{eqnarray}
(-1)^{w_{\rm 3d}}=\prod_{i}^8  
\frac{\sqrt{{\rm det}[w(\Gamma_i)]}}{{\rm Pf}[w(\Gamma_i)]}.
\label{eq:w3dPf}
\end{eqnarray}

This expression, Eq. (\ref{eq:w3dPf}), indicates that
$(-1)^{w_{\rm 3d}}$ depends only on the wave function at the special points
$\Gamma_i$ in the Brillouin zone, {\it i.e.}, the TRIMs.
On the other hand, a perturbation to remove a node may only affect the local
structure of the wave function near the gap node.
Therefore, Eq. (\ref{eq:w3dPf}) suggests that $(-1)^{w_{\rm
3d}}$ could be independent of the perturbation since the node is
not located at $\Gamma_i$ in general. 

To make the argument rigorous, we explicitly consider an odd-parity
superconductor.
For odd-parity superconducting states such as $\Delta_3$ and $\Delta_4$,
we have the following additional symmetry in the BdG Hamiltonian 
\begin{eqnarray}
\Pi {\cal H}({\bm k})\Pi^{\dagger}={\cal H}(-{\bm k}),
\label{eq:parity}
\end{eqnarray}
where $\Pi$ is given by the parity operator $P$ as
\begin{eqnarray}
\Pi=\left(
\begin{array}{cc}
P &0 \\
0 & -P^*
\end{array}
\right). 
\end{eqnarray}
Here we have used the fact that odd-parity superconductors satisfy
$P\Delta({\bm k})P*=-\Delta(-{\bm k})$.
In our BdG Hamiltonian \cite{HaoLee_B11}, $P=s_0\otimes \sigma_3$.
At the TRIM $\Gamma_i$, Eq. (\ref{eq:parity})
reduces to $[{\cal H}(\Gamma_i), \Pi]=0$, which means that the 
quasiparticle state
$|u_n(\Gamma_i)\rangle$ is simultaneously an eigenstate of $\Pi$, 
{\it i.e.},
$
\Pi |u_n(\Gamma_i)\rangle=\pi_n(\Gamma_i)|u_n(\Gamma_i)\rangle. 
$
Now let us consider the following matrix
\begin{eqnarray}
P_{mn}=\langle u_m({\bm k})|\Pi^{\dagger}\Theta |u_n^*({\bm k})\rangle,
\end{eqnarray}
which is connected to $w_{mn}$ at the TRIM $\Gamma_i$ via
\begin{eqnarray}
P_{mn}(\Gamma_i)=\pi_m(\Gamma_i)w_{mn}(\Gamma_i). 
\end{eqnarray}
Taking the Pfaffians on both sides yields 
\begin{eqnarray}
{\rm Pf}[P(\Gamma_i)]={\rm Pf}[w(\Gamma_i)] \prod_{m}\pi_{2m}(\Gamma_i), 
\end{eqnarray}
which leads to
\begin{eqnarray}
\prod_{i}\frac{\sqrt{\det [w(\Gamma_i)] } }{{\rm Pf}[w(\Gamma_i)]}
=\prod_{i}\frac{{\rm Pf}[P(\Gamma_i)]}{\sqrt{\det [P(\Gamma_i)}]}
\prod_m \pi_{2m}(\Gamma_i). 
\label{eq:PFpi}
\end{eqnarray}
We find here that the term 
${\rm Pf}[P(\Gamma_i)]/\sqrt{\det[P(\Gamma_i)]}$ gives rise to only a 
trivial factor. Indeed, by direct calculations, one obtains
\begin{eqnarray}
{\rm tr}A_i({\bm k})=-i \partial_i\ln {\rm Pf}[P({\bm k})]. 
\end{eqnarray}
Thus, by taking the ${\rm tr}A_i({\bm k})=0$ gauge [which is consistent with 
the relation ${\rm tr}F_{ij}({\bm k})=0$ obtained by the symmetries
(\ref{eq:time-reversal}) and (\ref{eq:parity})], one can see that 
${\rm Pf}[P({\bm k})]$ is independent of ${\bm k}$.  
Consequently, ${\rm Pf}[P(\Gamma_i)]/\sqrt{\det[P(\Gamma_i)]}$ is
factorized in Eq. (\ref{eq:PFpi}), which results in a trivial factor $1$.
Therefore, one obtains the final expression
\begin{eqnarray}
(-1)^{w_{\rm 3d}}=\prod_{i,m}\pi_{2m}(\Gamma_i). 
\end{eqnarray}

Now we can confirm explicitly that the mod-2 winding number is
independent of any weak perturbation $\delta$ we choose: The parity of
the wave function at the TRIM $\Gamma_i$ is determined locally by the
BdG equation at $\Gamma_i$, and hence it should not be affected by any
small perturbation away from the $\Gamma_i$ point. This means that we
have a unique value of the mod-2 winding number in the limit of
$\delta\rightarrow 0$, {\it i.e.}, the mod-2 winding number is
well-defined even in the presence of point nodes, although the
winding number $w_{\rm 3d}$ itself is not. Therefore, a
time-reversal-invariant superconductor with a pair (or pairs) of point
nodes can be robustly topological if the parity of the gap function is
odd.

In a weak paring state, we can evaluate $(-1)^{w_{\rm 3d}}$ rather easily.
In this case, $\prod_{i, m} \pi_{2m}(\Gamma_i)$ can be expressed in terms of
the energy dispersion of the system in the normal state \cite{Sato_B10, Sato09}, 
which leads to
\begin{eqnarray}
(-1)^{w_{\rm 3d}}=\prod_{i, m}{\rm sgn}[{\cal E}_{2m}(\Gamma_i)].
\end{eqnarray}  
Here ${\cal E}_{2m}({\bm k})$ is the normal-state energy dispersion
obtained as an eigenvalue of $\hat \xi({\bm k})$ [defined in Eq.
(\ref{hamiltonian})], and the Kramers degeneracy is taken into account as
${\cal E}_{2m}(\Gamma_i)={\cal E}_{2m+1}(\Gamma_i)$. In our model
Hamiltonian, we obtain $\prod_{i, m}{\rm sgn}[{\cal E}_{2m}(\Gamma_i)]=-1$. Thus,
all the odd-parity superconducting states $\Delta_2$, $\Delta_3$ and
$\Delta_4$ given in Table 1 have a non-trivial topological number
$(-1)^{w_{\rm 3d}}=-1$, indicating that all of them are topological
superconductors irrespective of the presence of the point nodes.

\clearpage

\begin{figure}
\begin{center}
\includegraphics[height=3.7cm]{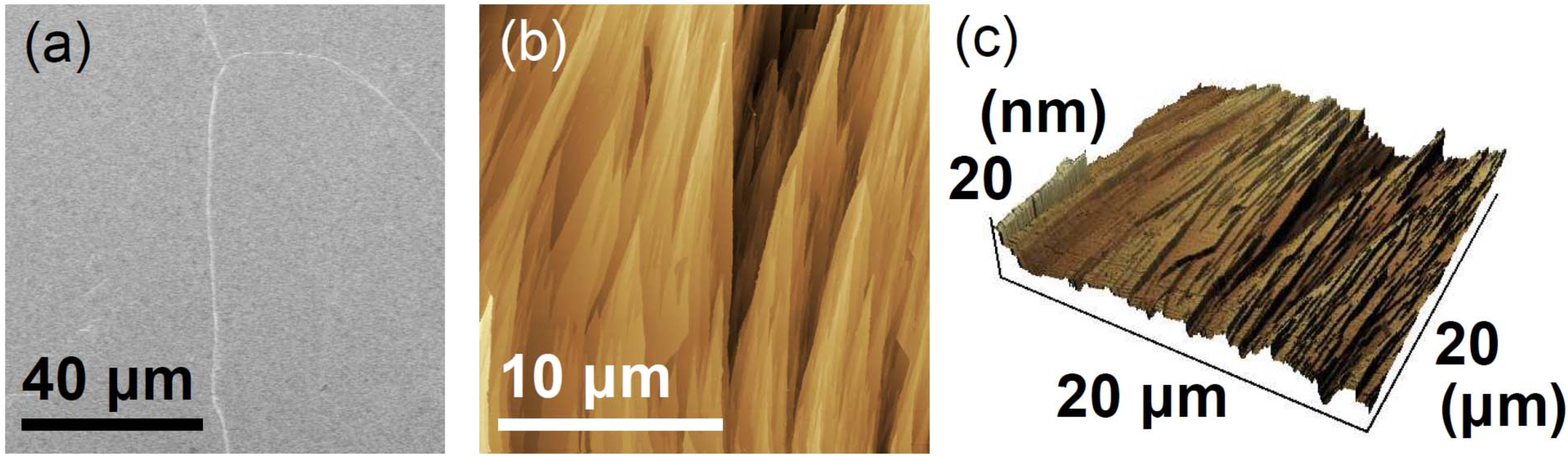}
\caption{Cleaved surface of Cu$_x$Bi$_2$Se$_3$. 
Scanning-electron microscope finds that the cleaved surface of Cu$_x$Bi$_2$Se$_3$ is
essentially flat (a). However, atomic-force microscope finds a lot
of nanometer-scale terraces in the ``flat" region (b). 3D
presentation of the data in panel {\bf b} is shown in panel (c),
where the vertical variation is very much exaggerated.
} 
\end{center}
\end{figure}

\begin{figure}
\begin{center}
\includegraphics[width=10cm]{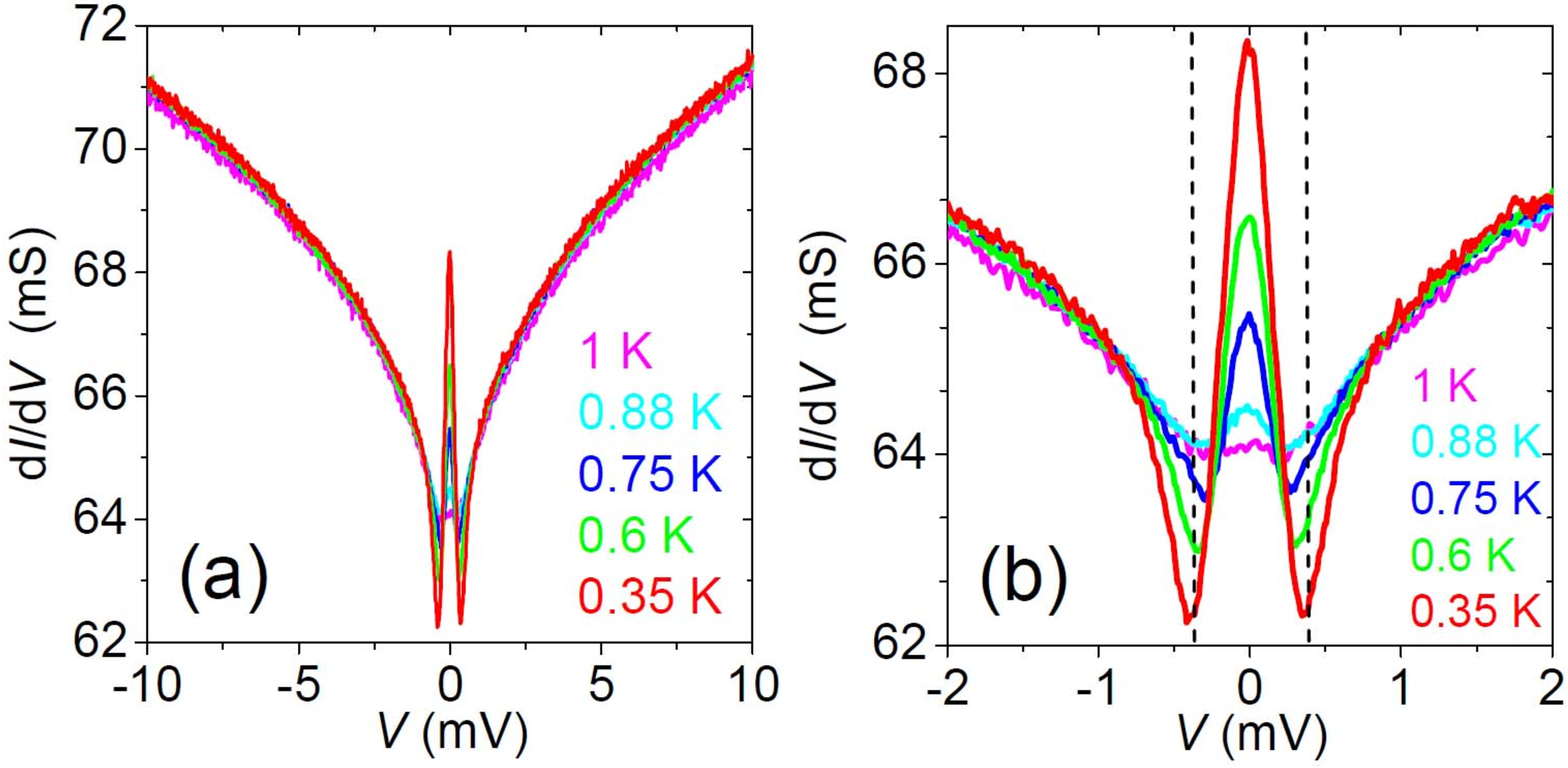}
\caption{Additional ZBCP data of Cu$_x$Bi$_2$Se$_3$. (a, b) 
Spectra showing the ZBCP in 0 T measured on
a sample different from that reported in the main text. (a) is
for a wide energy window, and (b) magnifies the low energy range.  
The dashed lines in {\bf b} indicate the energy position of the dips.
} 
\end{center}
\end{figure}

\begin{figure}
\begin{center}
\includegraphics[width=12cm]{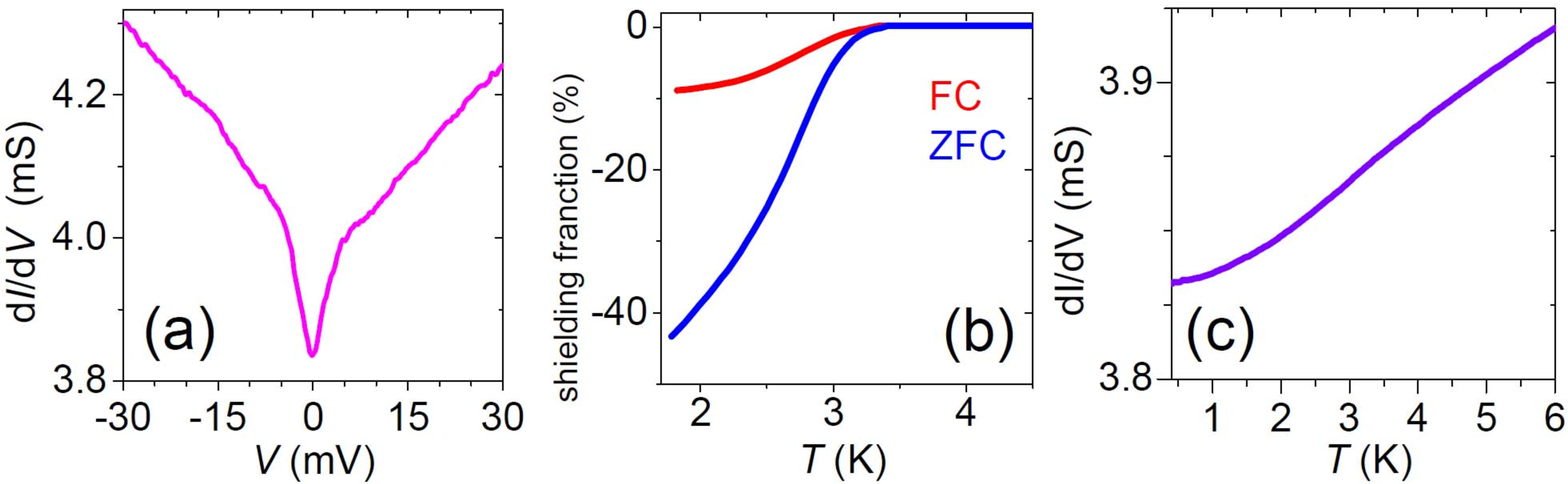}
\caption{An example of the point-contact spectrum at 0.35 K 
showing no feature associated with superconductivity (a), measured on a sample 
with the bulk onset $T_c$ of 3.2 K (b). The temperature dependence of 
the conductance of this point contact at zero bias showed no sign of 
superconductivity (c), suggesting that the portion of the sample beneath 
this point contact was non-superconducting.
} 
\end{center}
\end{figure}

\begin{figure}[t]
\begin{center}
\includegraphics[width=9cm]{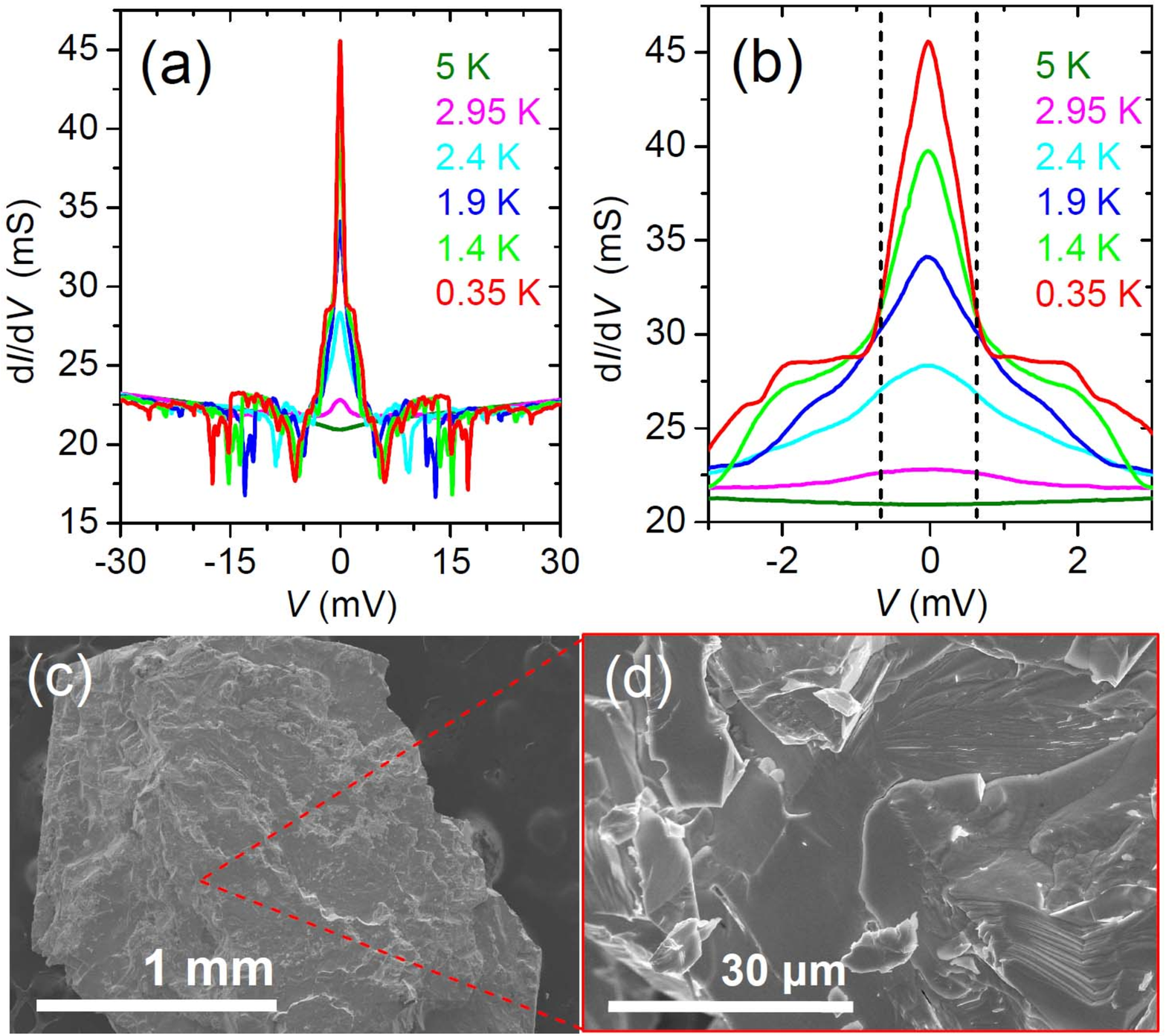}
\caption{Spectra of a point contact made on the as-intercalated surface 
of Cu$_x$Bi$_2$Se$_3$.
(a, b) $dI/dV$ vs bias voltage for 0.35 - 5 K measured in 0 T for a wide 
energy window is shown in (a), and (b) magnifies the data near zero energy; 
the vertical dashed lines indicate the sharp central part of the spectra in (b).
The as-intercalated surfaces are rough even under the optical microscope; 
(c) is a scanning-electron-microscope picture of a typical as-intercalated 
surface, and (d) is a magnified picture of a spot in (c).} 
\end{center}
\end{figure}

\vspace{-6mm}

\begin{figure}[htbp]
\begin{center}
\includegraphics[width=9cm]{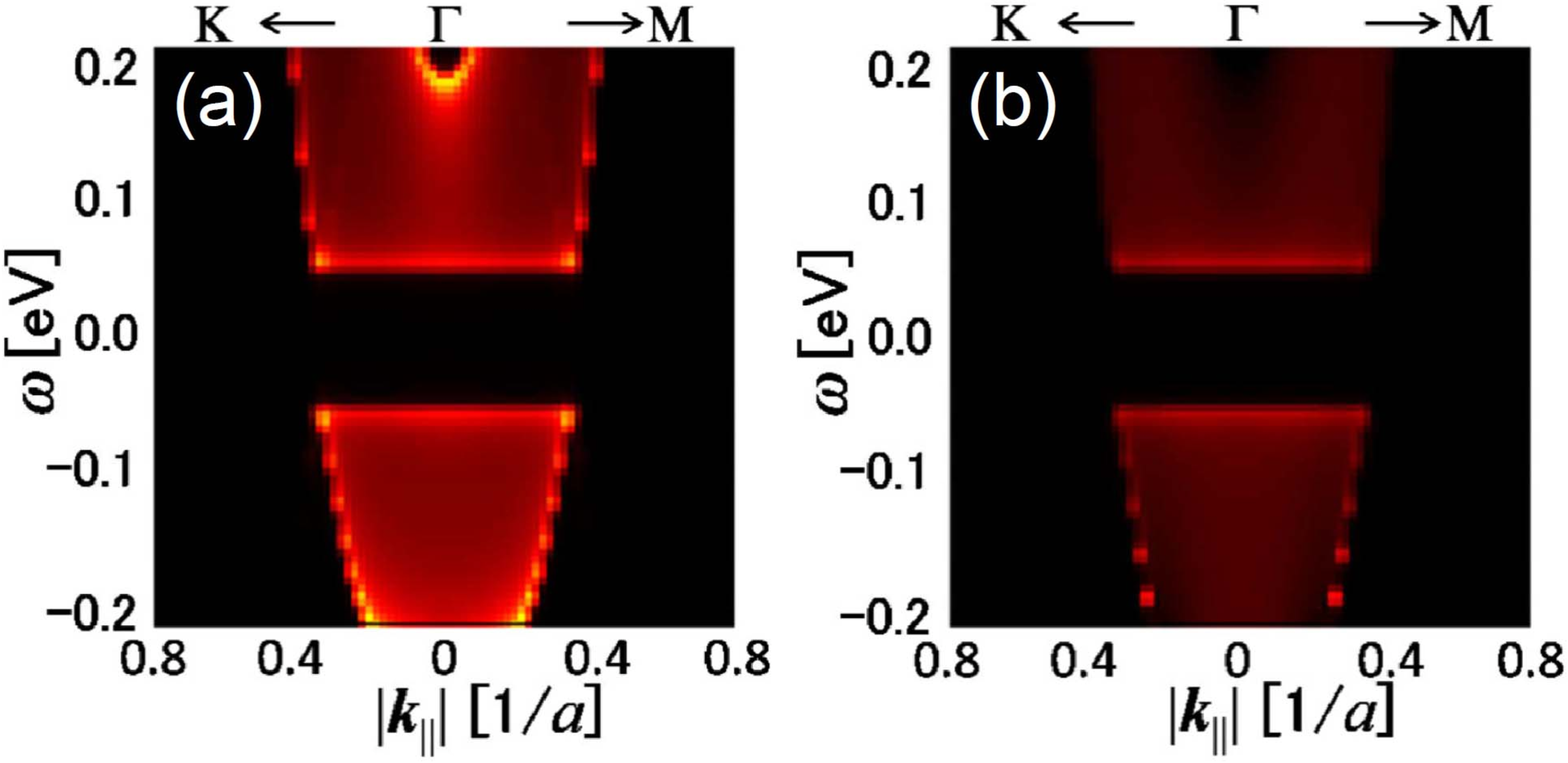}
\caption{
Spectral functions $A({\bm k}_\parallel,\omega)$ for the pair potential 
$\Delta_1$ ($\Delta^{11}_{\uparrow\downarrow}=-\Delta^{11}_{\downarrow\uparrow}
=\Delta^{22}_{\uparrow\downarrow}=-\Delta^{22}_{\downarrow\uparrow}$).
(a) bulk, (b) surface. }
\end{center}
\end{figure}

\vspace{-6mm}

\begin{figure}[htbp]
\begin{center}
\includegraphics[width=9cm]{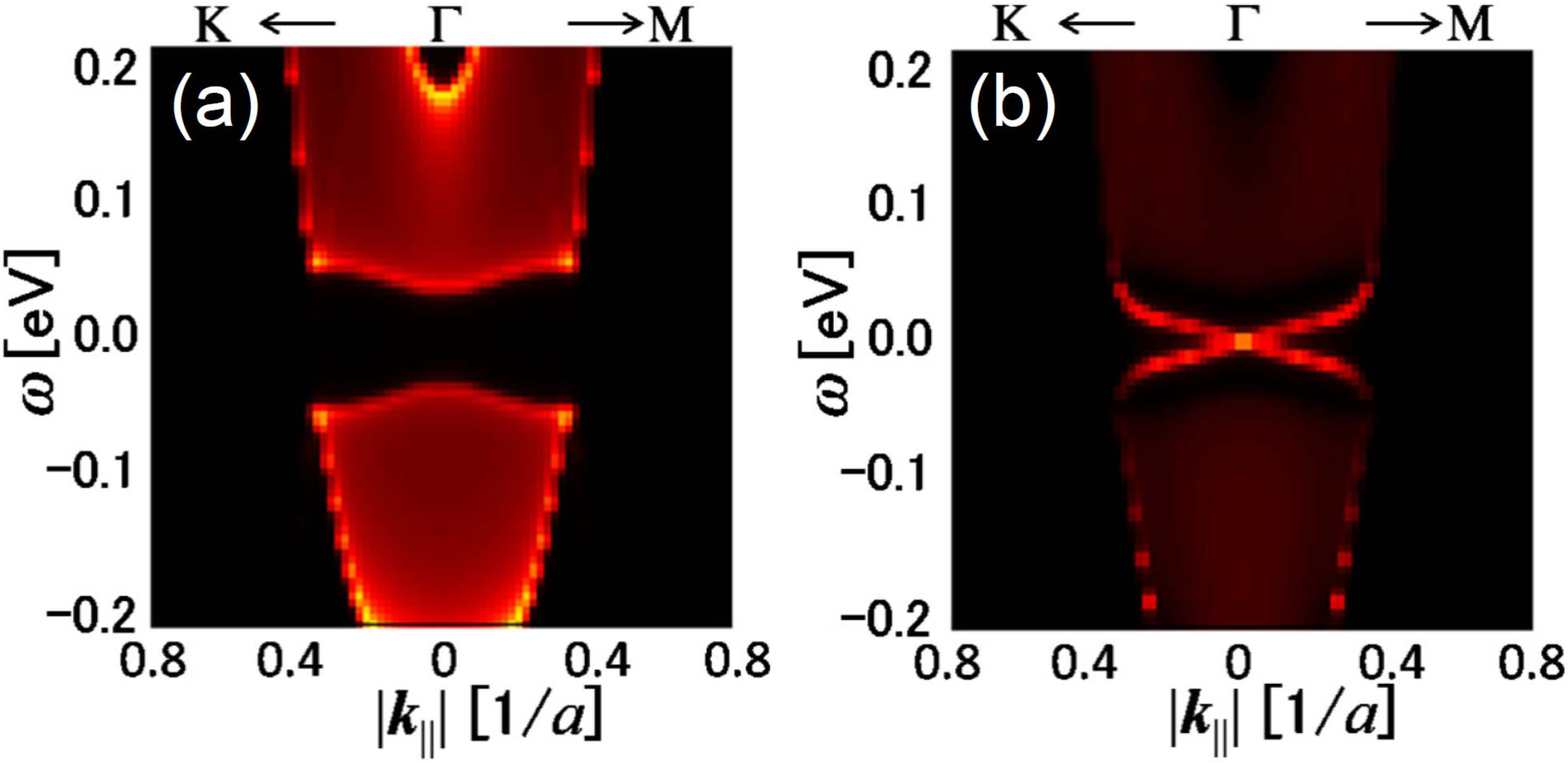}
\caption{
Spectral functions $A({\bm k}_\parallel,\omega)$ for the pair potential $\Delta_2$.
(a) bulk, (b) surface.}
\end{center}
\end{figure}

\vspace{-6mm}

\begin{figure}[htbp]
\begin{center}
\includegraphics[width=9cm]{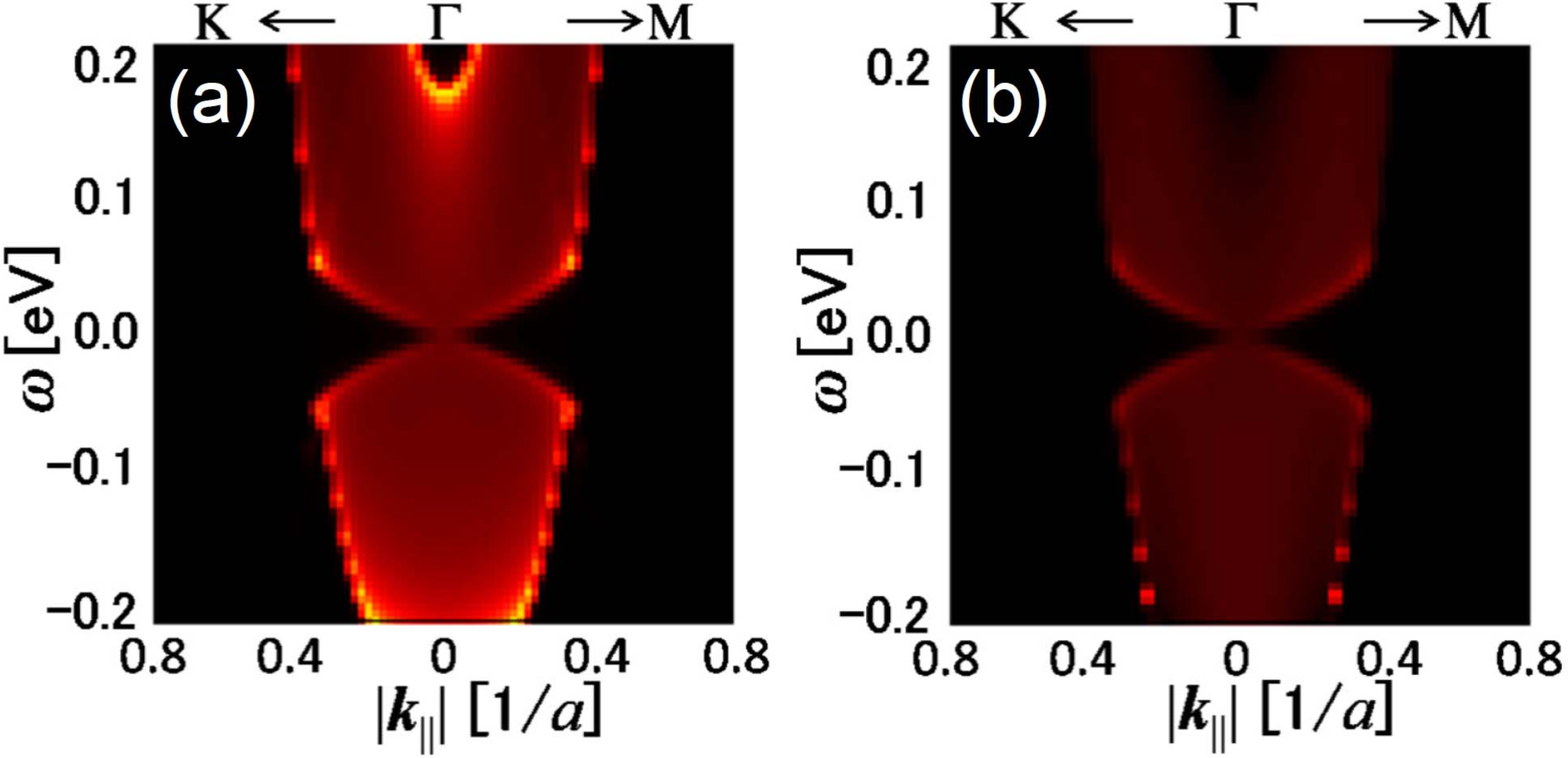}
\caption{
Spectral functions $A({\bm k}_\parallel,\omega)$ for the pair potential $\Delta_3$.
(a) bulk, (b) surface.}
\end{center}
\end{figure}

\vspace{-6mm}

\begin{figure}[htbp]
\begin{center}
\includegraphics[width=9cm]{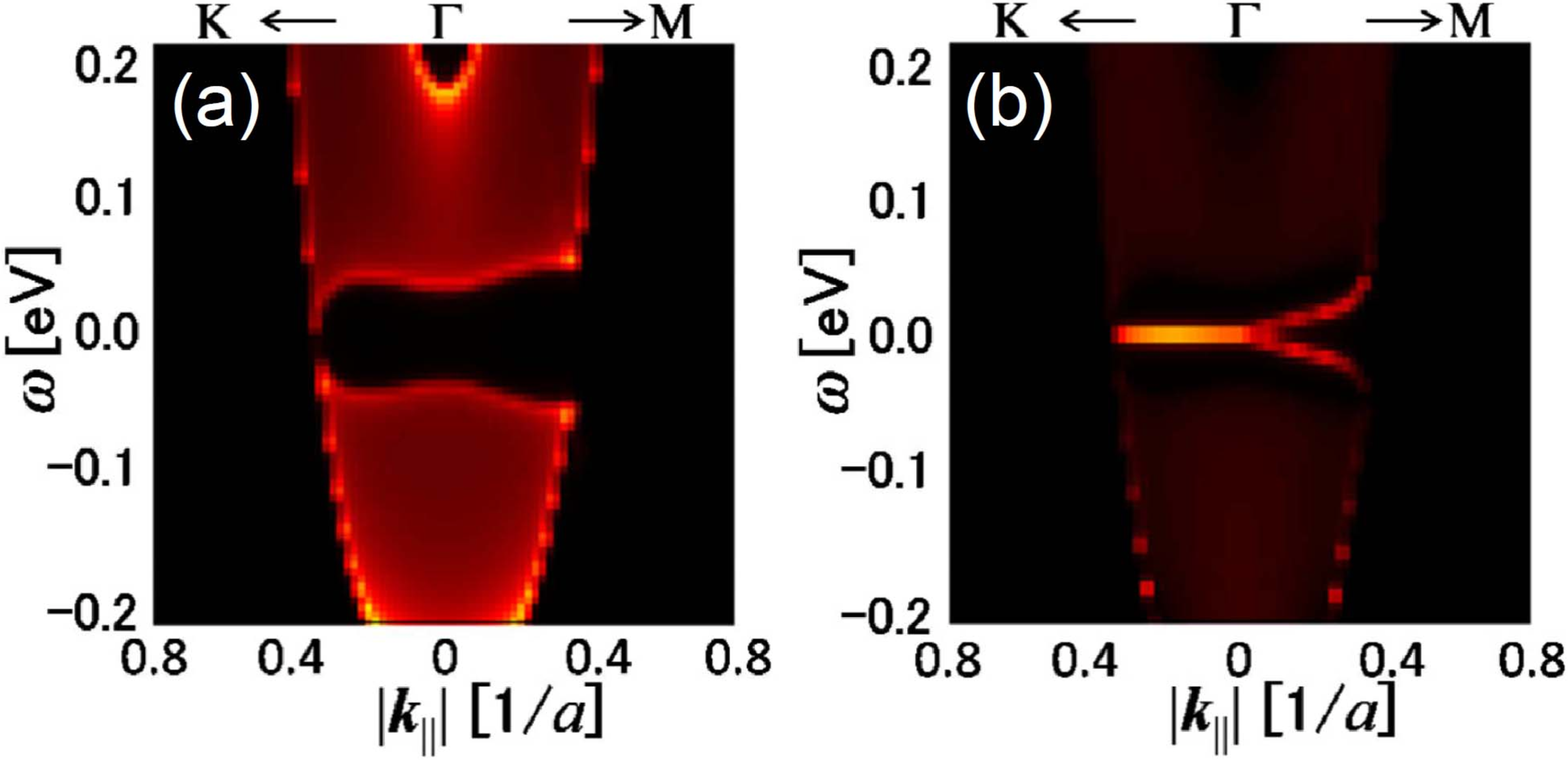}
\caption{
Spectral functions $A({\bm k}_\parallel,\omega)$ for the pair potential
$\Delta_4$ ($\Delta^{12}_{\uparrow\uparrow}=\Delta^{12}_{\downarrow\downarrow}
=-\Delta^{21}_{\uparrow\uparrow}=-\Delta^{21}_{\downarrow\downarrow}$).
(a) bulk, (b) surface.}
\end{center}
\end{figure}

\vspace{-6mm}

\begin{figure}[htbp]
\begin{center}
\includegraphics[width=8.5cm]{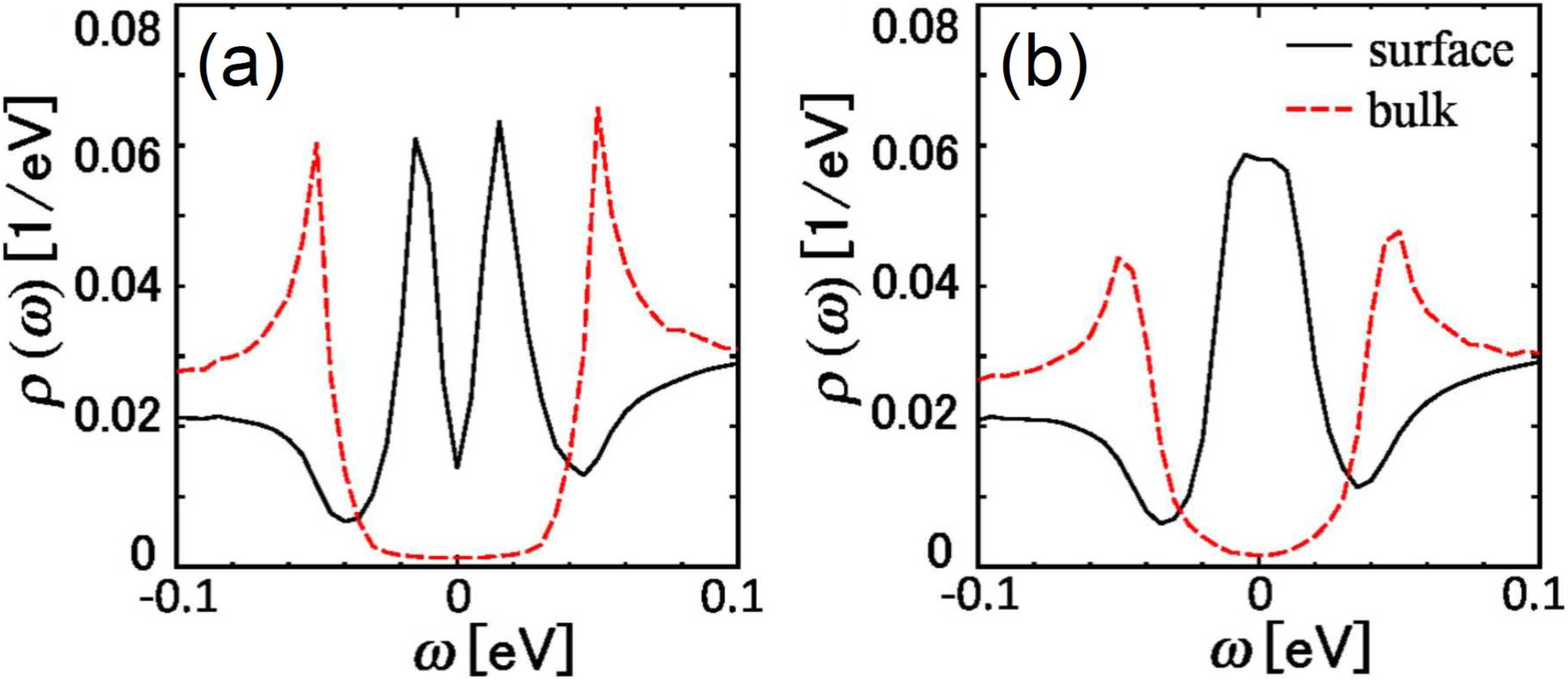}
\caption{
Calculated local density of states.  (a, b) LDOS at the bulk 
(dashed line) and the surface (solid line) for the 
pair potentials $\Delta_2$ [panel (a)] and $\Delta_4$ [panel (b)].}
\end{center}
\end{figure}

\vspace{-6mm}

\begin{figure}[htbp]
\begin{center}
\includegraphics[width=9.5cm]{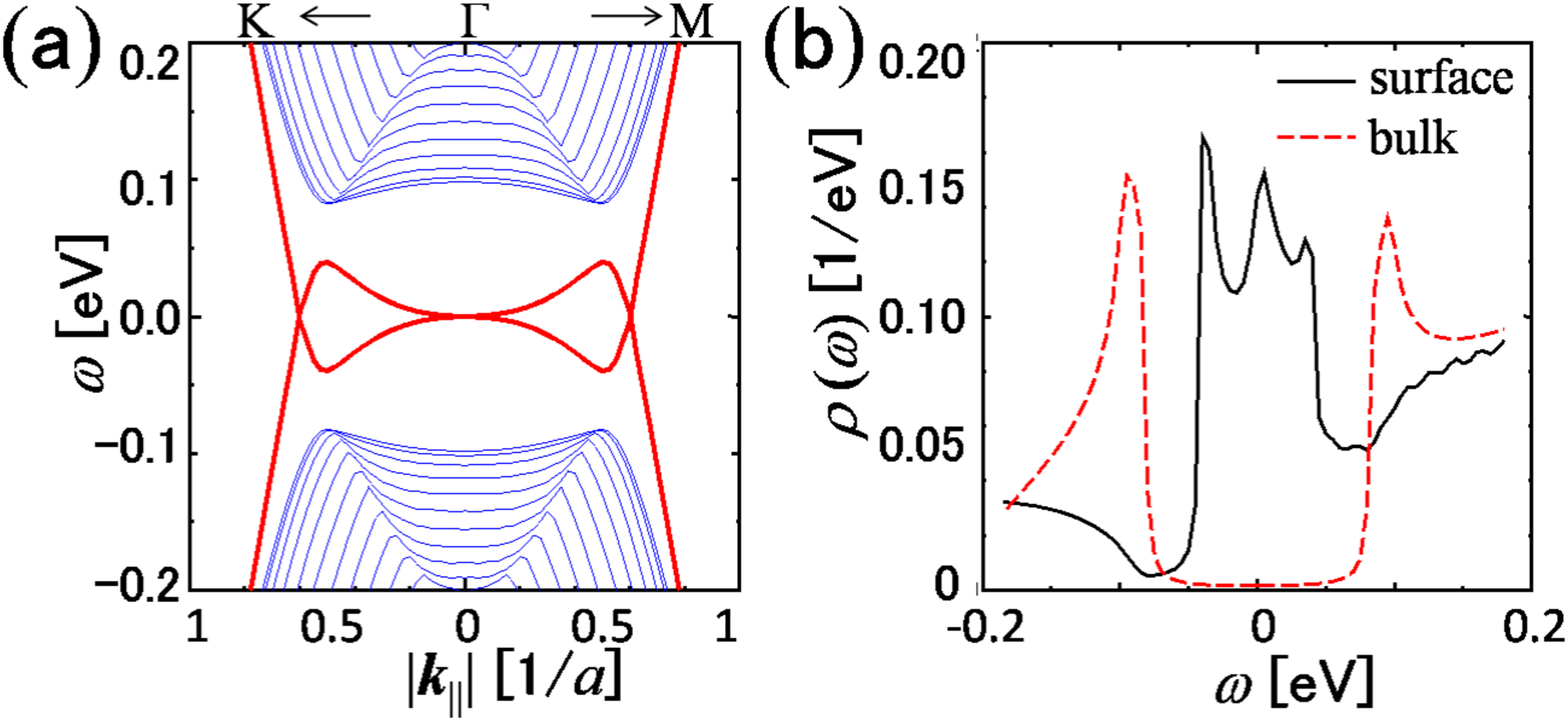}
\caption{
Calculation results for the pair potential 
$\Delta_2$ with 
$\bar{D_1} = \bar D_2 = 0$, $\bar A_1 = 1.0$, $\bar A_2 = 1.5$, $M_0 = -0.7$, 
$\bar B_1 = 0.5$,
$\bar B_2 = 0.75$, $\mu = 0.9$, and $\Delta(0) = 0.1$ (all in eV).
(a) Band dispersions 
for the bulk (blue) and the surface (red). (b) LDOS at the bulk 
(dashed line) and the surface (solid line).}
\end{center}
\end{figure}


\begin{thebibliography}{99}
\bibitem{note2}
The $g$ factor of Bi$_2$Se$_3$ is as large as 32 for $H \parallel C_3$ 
[H. K\"{o}hler and E. W\"{o}chner, Phys. Status Solidi B {\bf 67}, 665 (1975)],
which suggests that the Zeeman splitting for $S$ = 1/2 in 0.1 T would be 
$\sim$0.2 meV. This should be observable if the magnetic scattering is relevant.
\bibitem{note3}
With resistivity measurements, we confirmed that the SC state is completely
suppressed with the perpendicular magnetic field of 3 T (with the midpoint 
$H_{c2}$ of 1.7 T) at 0.35 K, consistent with the reported $H_{c2}(T)$ diagram 
\cite{MKR_L11}.
\end{thebibliography}
\end{document}